\documentclass{aa} 
\usepackage{graphicx}
\usepackage{natbib}
\usepackage{txfonts}
\usepackage{amsmath}
\usepackage{rotating}
\usepackage{graphics,wrapfig,times}
\usepackage[absolute]{textpos}

\begin{document}

\title{Six new doubly eclipsing quadruples in a 2+2 architecture}
 \author{Zasche,~P.~\inst{1},
 Ku\v{c}\'{a}kov\'{a},~H.~\inst{1,3,5,9},
 K\'{a}ra~J.~\inst{1,7},
 Merc~J.~\inst{1,8},
 Henzl,~Z.~\inst{2,3},
 Ma\v{s}ek~M.~\inst{3,4},
 \v{C}ervinka~L.~\inst{3,6},
 Dienstbier~V.\inst{1,3}}

 \institute{
  $^{1}$ Charles University, Faculty of Mathematics and Physics, Astronomical Institute, V~Hole\v{s}ovi\v{c}k\'ach 2, CZ-180~00, Praha 8, Czech Republic, \email{zasche@sirrah.troja.mff.cuni.cz}\\
  $^{2}$ Hv\v{e}zd\'arna Jaroslava Trnky ve Slan\'em, Nosa\v{c}ick\'a 1713, Slan\'y 1, 274 01, Czech Republic \\
  $^{3}$ Variable Star and Exoplanet Section, Czech Astronomical Society, Fri\v{c}ova 298, 251 65 Ond\v{r}ejov, Czech Republic \\
  $^{4}$ FZU - Institute of Physics of the Czech Academy of Sciences, Na Slovance 1999/2, CZ-182~21, Praha, Czech Republic\\
  $^{5}$ Research Centre for Theoretical Physics and Astrophysics, Institute of Physics, Silesian University in Opava, Bezru\v{c}ovo n\'am. 13, CZ-746 01, Opava, Czech Republic\\
  $^{6}$ Observatory Mlad\'a Boleslav, Svojs\'{\i}kova 1370, Mlad\'a Boleslav, Czech Republic \\
  $^{7}$ Department of Physics and Astronomy, University of Texas Rio Grande Valley, Brownsville, TX 78520, USA \\
  $^{8}$ Instituto de Astrof\'isica de Canarias, Calle Vía Láctea, s/n, E-38205 La Laguna, Tenerife, Spain \\
  $^{9}$ Astronomical Institute, Academy of Sciences, Fri\v{c}ova 298, 251 65, Ond\v{r}ejov, Czech Republic \\
 }

 \titlerunning{Six new 2+2 doubly eclipsing systems}
 \authorrunning{Zasche et al.}
 \date{Received \today; accepted ???}

\abstract{The study presents a confirmation of six quadruples with two sets of eclipses that have the 2+2 architecture. These so-called doubly eclipsing systems still present a quite rare group of stars. We collected all available photometric data and carried out a detailed analysis of them. In addition to the precise TESS photometry used to model the light curves for both inner eclipsing binaries, photometric survey data were also used, and more than 100 nights of our own dedicated observations were carried out. These were mainly used for the detection of the long-term evolution of orbital periods. Thanks to these data, we were able to derive the mutual orbits of the inner doubles via eclipse timing variations. The systems studied are: CzeV1254 (periods 0.715348 d + 0.362593 d, mutual period 15.6 yr); ASASSN-V J070838.27-171952.9 (4.300490 d + 3.767235 d, 4 yr); ASASSN-V J091951.17-593306.9 (2.668674 d + 14.342218 d, 0.95 yr); WISE J100820.07-731554.0 (0.368315 d + 7.945339 d, 3.8 yr); ASASSN-V J143536.01-721459.4 (7.353842 d + 6.368567 d, 27.4 yr); and Corot 310284765 (1.875418 d + 2.371126 d, 22.6 yr). Some of the binaries show significantly eccentric orbits. We also estimated their apsidal motion periods. With our six new systems, the number of confirmed orbits of 2+2 quadruples has been increased to 64 in total. }

 \maketitle

\section{Introduction}

Classical eclipsing binaries are sometimes presented as the foundations of modern stellar astrophysics. This is due to their role in 20th-century astrophysical research. Eclipsing binaries (hereafter EBs) are routinely used to precisely measure stellar radii and masses, as well as to calibrate the distance ladder  (nowadays even outside of our Galaxy), and stellar evolution theories via looking into stellar interiors (for a discussion of all of these aspects, see e.g. \citep{2012ocpd.conf...51S,2010A&A...519A..57C}).

In the current era of large-scale surveys and ultra-precise satellite data, it is not possible to perform a detailed analysis of every single EB in the sky, since their number now reaches more than 2 million \citep{2023A&A...674A..16M}. Therefore, it is essential to focus detailed research on some very specific types of EBs that are limited in number and that have high astrophysical importance, i.e. they offer us some unique insight into otherwise hard-to-detect processes and effects. By this we mean objects for which there are enough observational data for analysis, but that are still rare enough in the sky.

We are convinced that doubly eclipsing systems of the 2+2 architecture are clearly such objects. In total, among the known more than 2 million EBs, there are currently only about 1000 known candidates for doubly eclipsing systems. By a doubly eclipsing system candidate, we mean one point source in the sky, for which two different eclipsing periods are detectable in the data. We still mark them as candidates only, since we do not know anything about their true connection and gravitational coupling. Only after a proof of their mutual movement around a common barycenter is presented can we definitely say that these are real 2+2 quadruples.

However, it seems like the rate of optical doubles and blends of two close-by stars with eclipsing periods incorrectly attributed to only one star (i.e. false detections) is relatively low. Among the thousand known published doubly eclipsing systems, there are only several dozen proven optical doubles and blends (see e.g. \cite{2022MNRAS.511.3881F} and \cite{2024A&A...688A..41Z}). Therefore, one can ask why there are only about 60 definitely confirmed 2+2 quadruples among the order-of-magnitude-larger set of candidates with two known periods. We can see two different explanations for this apparent discrepancy. The first one is that the detection of the mutual orbit of the two doubles around a common barycenter is complicated and not always straightforward, and needs much more effort than the pure detection of the two eclipsing periods in the photometric data. The second reason could be the fact that we are still dealing with quite limited data. Mutual periods can be decades, centuries, or even thousands of years long, and our available time span of data is usually much shorter. Moreover, the orientation of the long orbit plays a role in the  successful detection of the orbit. On the other hand, the 2+2 quadruples are obviously more common in the stellar population compared to the 3+1 quadruples, and their parameters can give us useful hints concerning their formation mechanisms (see e.g. \cite{2021Univ....7..352T}).

Our presented work is a natural continuation of our long-term effort to detect and study these 2 + 2
quadruples. Without our contribution, the whole sample of the doubly eclipsing candidates would be
17\% smaller. Even more significant is the result of our previous publications on the number of proved
2+2 quadruples with derived mutual orbits. The total number of such systems would be about
65\% lower without our contribution. Therefore, we feel that the long-term effort of collecting the
ground-based data and analysing them in combination with the satellite ones is still useful for the
topic and brings us a non-negligible contribution of new systems.

\begin{table*}[t]
  \caption{Basic information about the systems.}  \label{systemsInfo}
  \scalebox{0.87}{
  \begin{tabular}{c c c c c c c}\\[-3mm]
\hline \hline\\[-3mm]
  Target name                          & UCAC4 name       &      TESS      &     RA     &     DE     & Mag$_{max}$ $^{\star}$ & Temperature/sp.type        \\
                                       &                  & identification &  [J2000.0] & [J2000.0]  &                        & information $^{\star\star}$ \\ \hline
 \object{CzeV1254}                     & UCAC4 732-033136 & TIC 260100848  & 04 11 07.5 & +56 22 32.9 & 13.30 (V) & $T_{eff} = 5731$ K \\
 \object{ASASSN-V J070838.27-171952.9} & UCAC4 364-023163 & TIC 148942773  & 07 08 38.3 & -17 19 52.9 & 13.41 (V) & $T_{eff} = 9390$ K \\
 \object{ASASSN-V J091951.17-593306.9} & UCAC4 153-026334 & TIC 386900081  & 09 19 51.2 & -59 33 06.9 & 13.83 (V) & $T_{eff} = 5783$ K \\
 \object{WISE J100820.07-731554.0}     & UCAC4 084-019866 & TIC 453459782  & 10 08 20.0 & -73 15 54.0 & 14.92 (V) & $T_{eff} = 5143$ K \\
 \object{ASASSN-V J143536.01-721459.4} & UCAC4 089-045921 & TIC 401924368  & 14 35 36.0 & -72 14 59.4 & 12.71 (V) & $T_{eff} = 5796$ K \\
 \object{Corot 310284765}              & UCAC4 422-081432 & TIC 44874302   & 18 33 51.3 & -05 39 23.5 & 15.28 (V) & sp B2IV$^{\sharp}$ \\
  \hline
\end{tabular}}\\
{\small Notes: $^\star$ Out-of-eclipse magnitude, $V_{mag}$, taken from UCAC4 \citep{2013AJ....145...44Z}. $^{\star\star}$ Effective temperature taken from the Gaia DR3 catalogue \citep{2023A&A...674A...1G}, or GAIA DR2 \citep{2018A&A...616A...1G}. $^{\sharp}$ \cite{2014yCat....102028C}. }
\end{table*}

\section{Target selection}

The targets selected for the current study were chosen from the set of candidates among the published systems. Most of them are from our recent paper on TESS discoveries \citep{2022A&A...664A..96Z}. One was taken from an older paper dealing with the Corot data \citep{2017MNRAS.471.1230H}. The selected stars are a mixture of northern and southern targets. Their spectral types or masses also did not play a role in our selection process. The only criterion used was the detection of two eclipsing periods in the available data and the non-existence of any detailed analysis of the star. From this point of view, our present paper can be considered as a first confirmation of these stars as proven quadruples, i.e. with their mutual orbit firstly derived. A summary of information about the six targets is presented in Table \ref{systemsInfo}. One can see there that the stars are relatively faint targets, and we have only very limited information in our hands at present (i.e. no detailed spectroscopy).

The detection of the systems as doubly eclipsing was originally done by \citep{2022A&A...664A..96Z}, who easily plotted the TESS data where the additional eclipses are directly visible by the naked eye. However, the derivation of the proper period of the secondary pair was done by subtracting the dominant eclipsing pair, and on the residuals the period was derived more robustly over several TESS sectors of data.

The selected systems analysed in our study seem to be only a drop in the ocean, and not exceptional among the other 2+2 quadruples that have already been proven. A comparison with the other systems is summarised below in Section \ref{Conclusions}. What can be stated already is the fact that we are still dealing with quite a low number of contact binaries as the inner pairs among these doubly eclipsing systems (see Figure 2 in our recent paper \citealt{2022A&A...664A..96Z}). Only about one quarter of the currently proved doubly eclipsing 2+2 quadruples contain at least one inner binary showing a contact-shaped light curve (hereafter LC). Even in our present paper, only one system (the first one, CzeV1254) shows such a LC. However, this deficiency is still probably just a consequence of the problematic analysis of these contact LCs and their disentangling, rather than an indication that there is a lack of such systems in the stellar population. As was already stated above, our contribution to the topic with our six new systems represents 10\% of the total number of known 2+2 doubly eclipsing quadruples. With hundreds of them in the future, we can hope to find some clues concerning their formation mechanism, for example (see e.g. \cite{2021Univ....7..352T} discussing different formation channels between 2+2 and 3+1 quadruples).

\section{Data handling and fitting} \label{fiting}

The TESS data \citep{2015JATIS...1a4003R} were the most important source of photometric information for our systems. This is due to their accuracy as well as their cadence. We extracted the photometry from the TESS archive using the software {\sc lightkurve}  \citep{2018ascl.soft12013L}. The TESS data were used for the analysis of LCs of both inner binaries (except for the last system).

For the fitting of the LCs, the software {\sc PHOEBE} \citep{2005ApJ...628..426P} was used. It uses the standard Wilson-Devinney algorithm \citep{1971ApJ...166..605W} with its later modifications. When dealing only with photometry, several important input parameters have to be fixed or estimated elsewhere.

The orbital period of a binary has to be derived outside of {\sc PHOEBE}. This preliminary period estimation was done on the best TESS sector of data. After that, the period was step-by-step fixed for some short time interval, for example for one TESS sector, as it is apparently changing continuously as the binary revolves around a common barycenter with the other binary in the system. Such an approach is well substantiated due to the fact that the duration of one TESS sector of data (i.e. 27 days) is always an order of magnitude shorter than the period variation on the long orbit. Therefore, it produces errors in the times of minima at the level of 10$^{-5}$ to 10$^{-4}$ day at maximum, i.e. an order of magnitude smaller than typical uncertainties of the times of minima derivation precision. For the mass ratio of the particular binary, we started with the assumption of equal masses, i.e. $q=1.0$, but later this parameter was also released and fitted for systems with larger out-of-eclipse variations. 

For the whole analysis, we followed our step-by-step iterative procedure. The same method was already used for our previous studies on the topic (see e.g. \cite{2024A&A...687A...6Z,2023A&A...675A.113Z}). It means that we first performed a preliminary fit of the most pronounced pair, A. Then this fit was subtracted from the complete (A+B) photometry and the resulting pair, B, was fitted as well. This preliminary fit of pair B was again subtracted from the original combined photometry and only pair A was fitted again. This procedure was repeated several times until we obtained the final acceptable results for both pairs.

Together with the fitting of the LCs, the periods of both pairs were computed with a higher precision as well. This step was also done iteratively several times. We used our method of AFP (automatic fitting procedure, introduced in \citealt{2014A&A...572A..71Z}), which deals with the LC template and provides us with the times of eclipses. From the times of eclipses, the better period for the particular pair can be derived. With such an improved orbital period, a better LC fit can be obtained. We implemented this step-by-step procedure for both inner eclipsing pairs, A and B, until we got the final results for both pairs. For this AFP method, the orbital period was also fixed at its actual value (a linear fit from the long-term period variations) for particularly short intervals of data, i.e. hundred(s) of days typically.

We also have to be aware of the fact that the TESS data may in principle have different levels of light contamination in different sectors. This aspect of the whole TESS photometry was taken into account when fitting the LCs of both pairs. This means that the third light values of both A and B pairs were fitted independently as well. Therefore, we consider the third light value to be a free parameter for both pairs, trying to estimate the light contributions of both A and B pairs. However, due to varying values in between different sectors, this can only give us a rough estimate. The values presented below in Table \ref{TabLC} give only one value for the TESS sector with the best data (the one with the shortest cadence of data). In principle, this value can be higher than 100\% for pairs A and B in total, which can indicate additional light from some close-by components that are not necessarily connected with the system itself, for example. We assumed that the value of the third light is constant (i.e. comes from a non-variable star) during at least one TESS sector of data, for instance.

The proper orbit of the two doubles around a common barycenter (and hence the proof of their 2+2 quadruple nature) can be derived from the long-term ETVs (eclipse timing variations). Both ETV curves of the A and B pairs should behave in the opposite manner.

Both the older (and less precise) photometry and the new dedicated ground-based observations from our observers can help us greatly in this regard, since the TESS photometry itself covers only a small part of the long orbit. Including these observations, our time baselines improved to much longer time intervals of about 20 years in some cases. The new observations were obtained mainly on these observation sites:
\begin{itemize}
    \item The Danish 1.54-m telescope on La Silla, Chile, remotely operated, equipped with a CCD camera, and standard filters,
    \item The 65-cm telescope at Ond\v{r}ejov observatory in Czech Republic, using a MII G2-3200 CCD camera,
    \item The 30-cm telescope at a private observatory in Velt\v{e}\v{z}e u Loun, Czech Republic, equipped with a MII G2-8300 CCD camera,
    \item The 20-cm telescope at a private observatory in Doln\'{\i} Bousov, Czech Republic,
    \item Two FRAM telescopes of 30 and 25 cm, located on La Palma \citep{2019ICRC...36..769P}, and in Argentina \citep{2021JInst..16P6027A},
    \item The 30-cm ROST telescope of the Variable Star and Exoplanet Section of the Czech Astronomical Society located at Ond\v{r}ejov observatory in the Czech Republic, equipped with a MII G4-16000 CCD camera.
\end{itemize}

For the analysis of the period variations of both pairs, we used a standard least-squares method of fitting the data with the theoretical curve. By this curve we mean a standard LITE (light-time effect, or also named R\"omer delay) curve describing the orbit around a barycenter with the distant body (see e.g. \cite{1990BAICz..41..231M}, or \cite{2016MNRAS.455.4136B}):
\begin{equation} \,\,\,\,ETV_{LITE} = \frac{A_{LITE}} {\sqrt{1-e^2\cos^2\omega}} \left[ {{ \frac{(1\!-\!e^2) \!\cdot \sin(\nu + \omega)}{1+e \cos\nu} }} \right].\end{equation}
However, for some tighter systems where the outer orbit of the two binaries is close enough to the inner two orbits (i.e. for orbits with periods of $p_{A-B} \lesssim 1$yr) one also has to take into account the dynamical interaction between the orbits. We used an approach that has been presented for this so-called ‘physical delay’, for example by \cite{2013ApJ...768...33R}:
\begin{multline}
    \,\,\,\,ETV_{phys} =  A_{phys} ( (2\cdot\cos^2 i_m - 2/3) \cdot (\nu + e\sin \nu - M) +\\(1-\cos^2 i_m) \cdot (\sin(2\nu-2v_m) + e\cdot\sin(\nu-2v_m) + \frac{e}{3} \sin(3\nu-2v_m)))  .
\end{multline}
This means that for the more compact systems, both of these effects should be taken into account together and added. In these two equations, (1) and (2), there are the following symbols: $A$ stands for the amplitudes of both effects, $e$ is the eccentricity of the outer orbit and $\omega$ its argument of periastron, and $\nu$ is the true anomaly on the long orbit, while $M$ is the mean anomaly, $v_m$ stands for the orientation of the periapse of the outer orbit to the inner binary orbit, and $i_m$ is the mutual inclination of the inner binary and the outer orbit. The amplitudes of both these effects depend on the semi-major axes of the orbits, or the masses of the individual components. At least in principle, all these computable quantities should be derived from good data (which means data that are well distributed in time and that have an adequate accuracy).

 \section{Individual systems}

In the following sections, we focus on each system from our sample one by one, emphasising their differences and individual aspects of the analysis.

\subsection{CzeV1254}

The very first system is CzeV1254 (= UCAC4 732-033136, $V=13.3$ mag out of eclipse). The star was first discovered as a classical EB. Later, in \cite{2022A&A...664A..96Z}, it was also classified as a doubly eclipsing candidate due to the discovery of two eclipsing periods (of EA+EW type): 0.715419 + 0.362593 days. Both EBs show rather deep eclipses of about 0.1 mag; however, no detailed analysis of the system has been published to date. According to the GAIA DR3 temperature estimate, the system is probably of G-dwarf type, being about 565 pc distant from the Sun (GAIA DR3).

 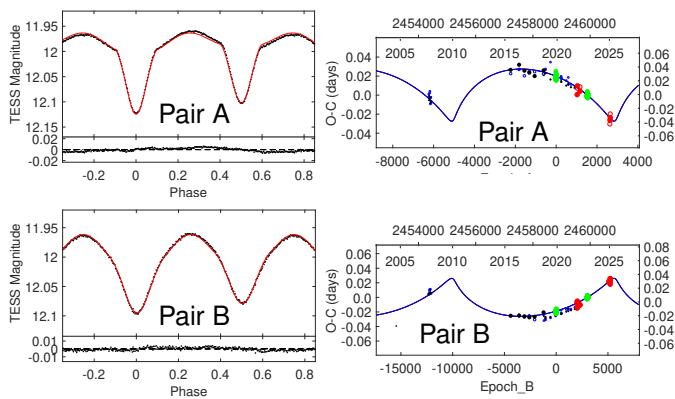
\begin{figure}
 \centering
 \begin{picture}(380,145)
  \put(1,75){\includegraphics[width=0.22\textwidth]{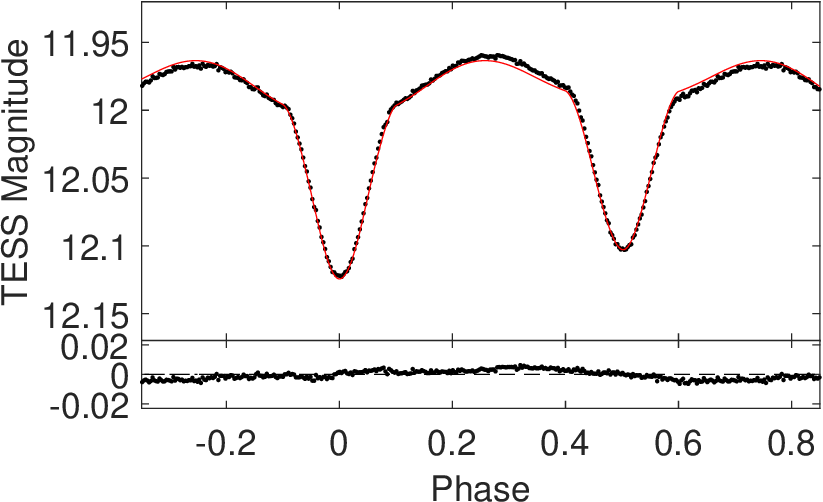}}
  \put(1,0){\includegraphics[width=0.22\textwidth]{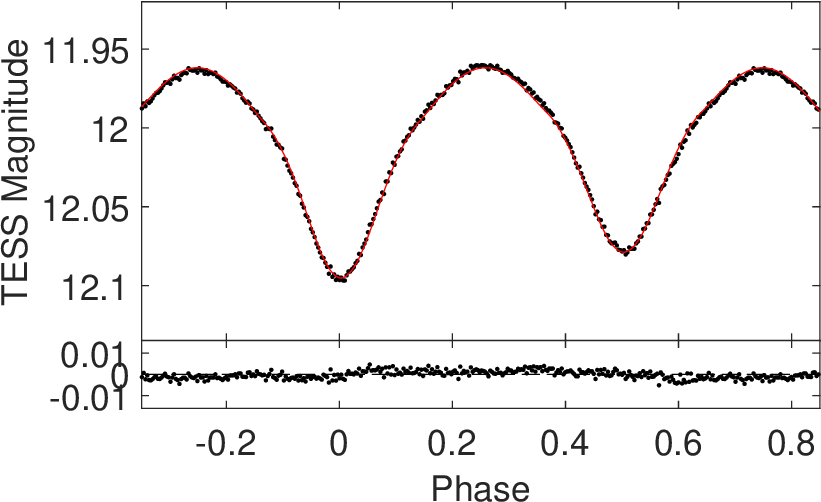}}
  \put(120,79){\includegraphics[width=0.25\textwidth]{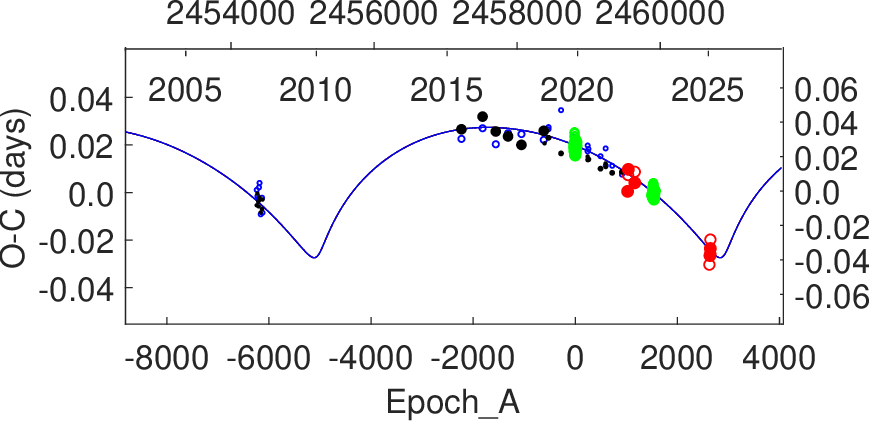}}
  \put(120,1){\includegraphics[width=0.25\textwidth]{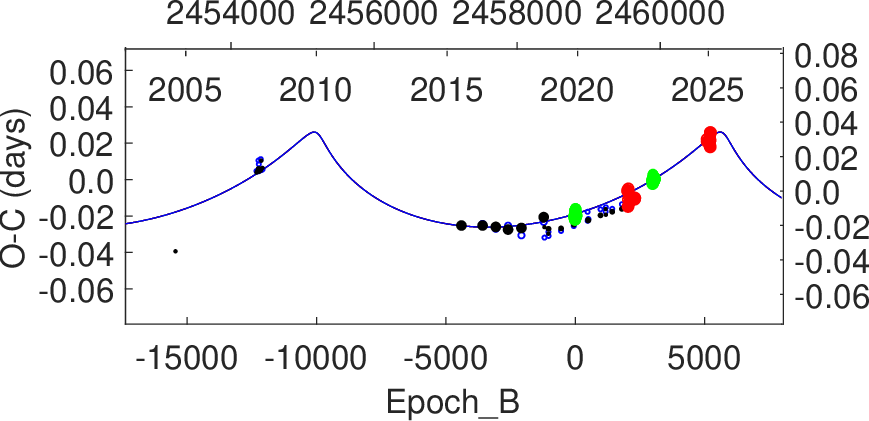}}
   \put(52,103){ { \textsf{Pair A}}}
   \put(52,27){ { \textsf{Pair B}}}
   \put(172,96){ { \textsf{Pair A}}}
   \put(150,20){ { \textsf{Pair B}}}
  \end{picture}
  \caption{Analysis of the CzeV1254 system. Left-hand side: Light curves of both disentangled pairs, A and B, based on the TESS data (residuals given in the bottom plots). Right figures: ETV diagrams of both pairs as they revolve around a common barycenter. Full dots stand for primary eclipses, and open circles for the secondary ones. Green symbols represent the TESS data, and red ones our new dedicated observations. The remaining symbols are the other photometric surveys.}
  \label{Fig_Czev1254}
 \end{figure}

Of all the available TESS sectors, we selected sector 59 for further analysis. The two overlapping photometric periods are clearly seen there. We were able to separate the two signals using the methods described in Section \ref{fiting}. Despite the fact that both pairs have rather short periods, we were able to disentangle them and perform the independent LC analysis for both pairs. The {\sc PHOEBE} fitting led to the LC parameters given in Table \ref{TabLC} and the fit presented in Fig. \ref{Fig_Czev1254}. One can see there that the light contribution of both pairs is roughly the same. It means both binaries should also be of a similar mass (using the standard and easiest assumption of the main-sequence mass-luminosity relation). The slight asymmetry of a pair A light curve is probably caused by some photospheric spots, but we have not tried to fit this.

For the study of period changes in both pairs, long-term evolution, and the detection of mutual movement of both pairs around a barycenter, we collected all available data for the star. This meant that apart from the new TESS data and our new dedicated observations, we also used older ASAS-SN data \citep{2018MNRAS.477.3145J,2017PASP..129j4502K}, SWASP \citep{2006PASP..118.1407P}, and ZTF \citep{2019PASP..131a8003M}. Thanks to these data, the interval covered with data spread to 20 years. Therefore, our fit presented in Fig. \ref{Fig_Czev1254} with its periodicity of about 15.6 years (see Table \ref{TabETV}) is now covered. It is still possible that, due to rapid period change near the periastron passage, the period is a bit longer. Therefore, new observations in the upcoming seasons would still be useful to obtain. A slight asymmetry of the LC as is presented in the residuals in Fig. \ref{Fig_Czev1254} could also cause an improper derivation of the time of the mid-eclipse, or even additional apparent period variation in the ETV diagrams (see e.g. our study on the topic published in \citealt{2011IBVS.5991....1Z}). However, in this case, only a slightly worse time of minimum derivation is caused, shown by a larger scatter in the ETV diagrams. The overall picture of our fit remains practically the same, since the fit is derived through the opposite behaviour of A and B pairs. A similar evolution of asymmetries for both pairs in time is highly improbable.

The comparable amplitudes of the A and B pairs indicate similar masses of both pairs, which is in agreement with the third light ratios derived from the LC analysis as presented in Table \ref{TabLC}. Due to its relatively small distance from the Sun, the value of the predicted angular separation of the double in the sky of about 30~mas is quite favourable. This is well within the limits for modern interferometers. However, more problematic is its relatively low brightness for such a technique.

\subsection{ASASSN-V J070838.27-171952.9}

The second system analysed was ASASSN-V J070838.27-171952.9 (= TIC 148942773, $V=13.41$ mag out of eclipse). This eclipsing system in the Canis Majoris constellation was first discovered as an EB by the ASAS-SN survey, which detected its prominent 4.3-day eclipses. However, with precise TESS data we were able to detect a secondary periodicity of about 3.77 days. Both of  these eclipse-shaped curves indicate well-detached systems for both A and B binaries. No other detailed analysis exists and no spectroscopy is available for this star.

 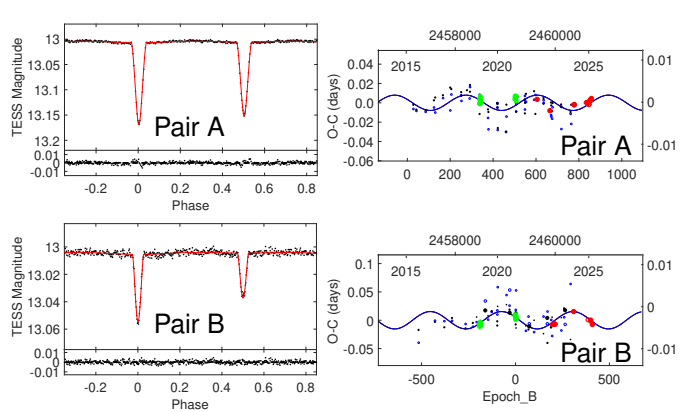
\begin{figure}
 \centering
 \begin{picture}(380,150)
  \put(1,75){\includegraphics[width=0.22\textwidth]{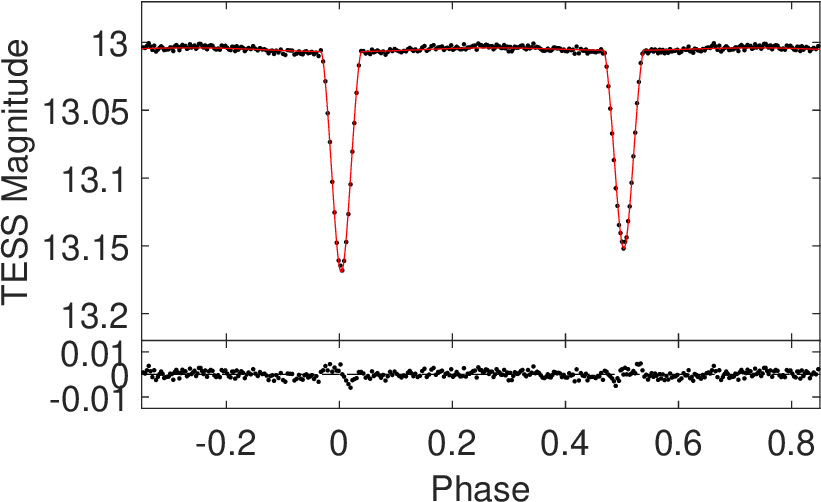}}
  \put(1,0){\includegraphics[width=0.22\textwidth]{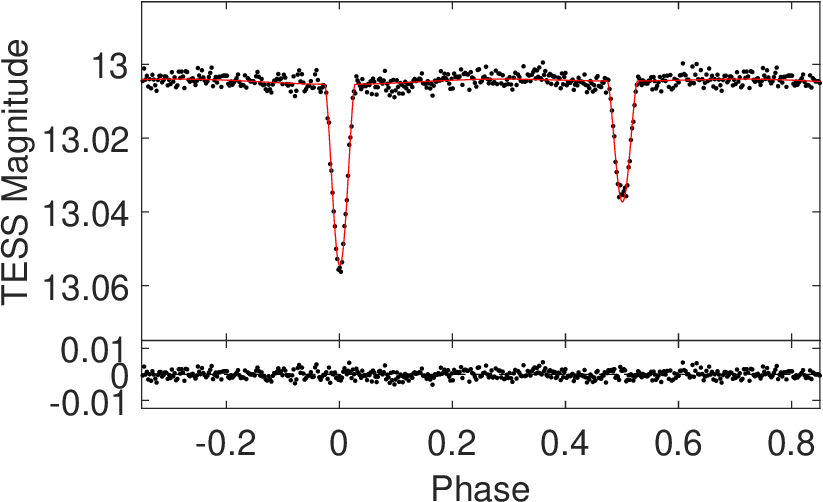}}
  \put(120,79){\includegraphics[width=0.25\textwidth]{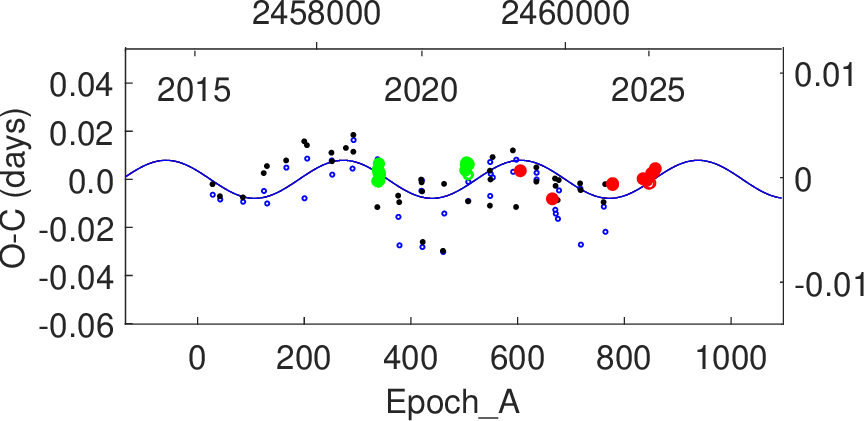}}
  \put(120,2){\includegraphics[width=0.25\textwidth]{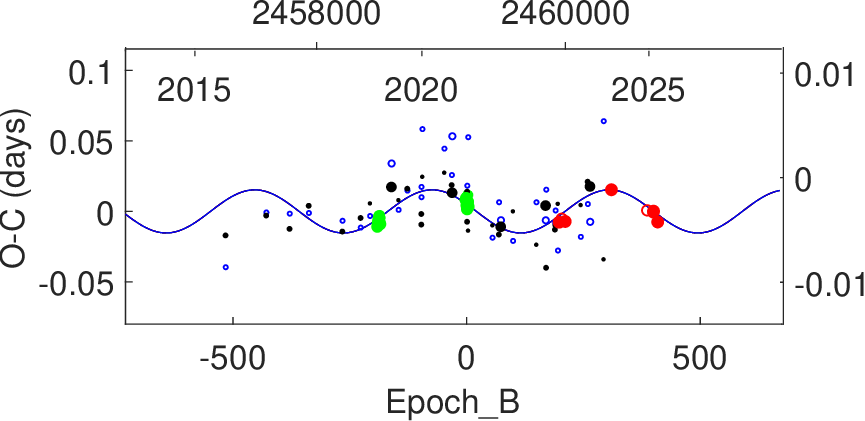}}
   \put(52,103){ {\textsf{Pair A}}}
   \put(52,29){ {\textsf{Pair B}}}
   \put(205,96){ {\textsf{Pair A}}}
   \put(205,18){ {\textsf{Pair B}}}
  \end{picture}
  \caption{Analysis of the ASASSN-V J070838.27-171952.9 system. The description of the plots and colour coding for individual symbols are the same as in Figure \ref{Fig_Czev1254}.}
  \label{Fig_asas070838}
 \end{figure}

We followed more or less the same approach as for the first system. The TESS data were analysed and the LC fit was derived (using TESS sector 33). The parameters of this {\sc PHOEBE} fitting are given in Table \ref{TabLC}. Despite the detached configuration, both inner pairs are circular (the small eccentricity value of pair A is still rather uncertain, see the residuals in Fig.\ref{Fig_asas070838}). Due to only very small out-of-eclipse variation, we did not try to fit the mass ratio of either of the pairs and let this parameter stay fixed to its suggested value of equal mass, $q=1.0$. Pair B seems to be the less luminous one (as was expected from a much shallower eclipse depth). The LC fits of pairs A and B are plotted in Fig.\ref{Fig_asas070838}.

Collecting all older photometric data, we tried to detect the long-term period modulation of both pairs. We used the following sources for this system: ASAS-SN, ZTF, Atlas \citep{2018AJ....156..241H}, and our own data. The result of our fitting is plotted in Fig. \ref{Fig_asas070838}, and the results are written in Table \ref{TabETV}. Due to the rather poor quality of the data and the only very small amplitude of ETV variations in both pairs, we used only a simplified approach of a circular outer orbit with zero eccentricity. A much larger number of more precise observations would be needed for a better characterisation of this mutual orbit.

\subsection{ASASSN-V J091951.17-593306.9}

The third star analysed is ASASSN-V J091951.17-593306.9 (= TIC 386900081, $V=13.83$ mag out of eclipse). This star was originally also classified as a standard EB with its orbital period of about 2.669~days, based on the ASAS-SN data \citep{2020MNRAS.491...13J}. However, thanks to new precise TESS data we were able \citep{2022A&A...664A..96Z} to detect an additional secondary eclipsing period of about 14~days. No other detailed analysis for this star was published to date.

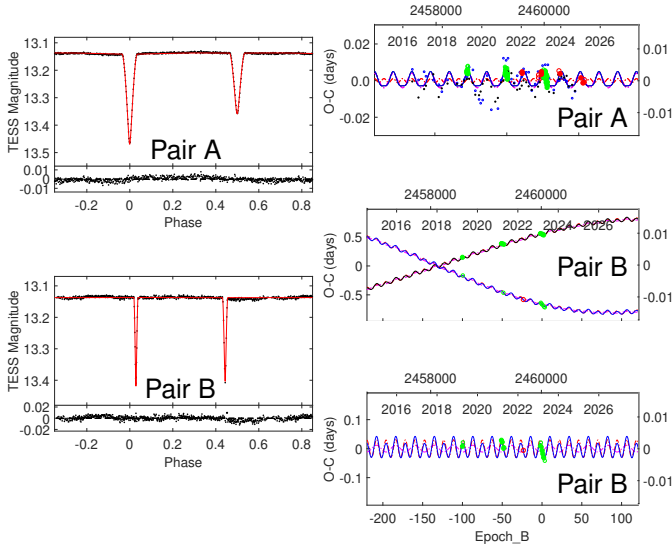
\begin{figure}
 \centering
 \begin{picture}(380,200)
  \put(1,120){\includegraphics[width=0.22\textwidth]{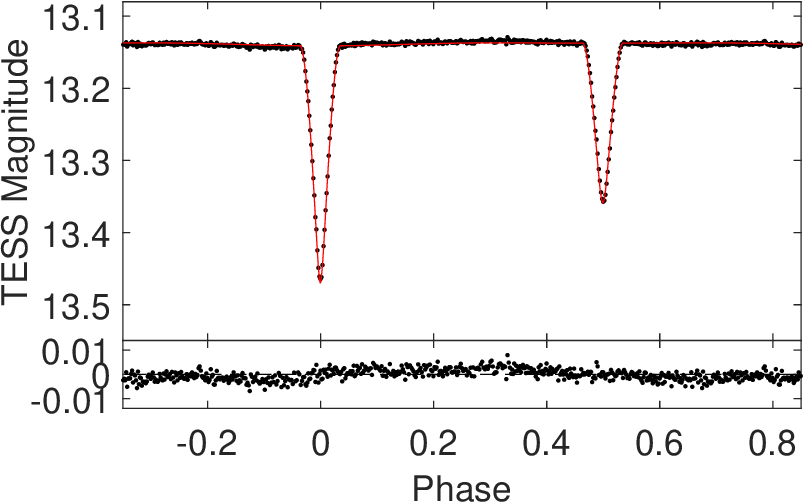}}
  \put(1,30){\includegraphics[width=0.22\textwidth]{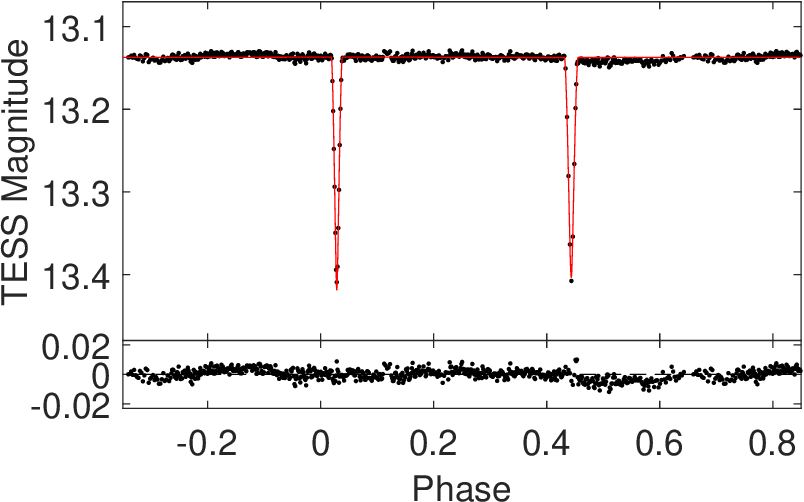}}
  \put(120,140){\includegraphics[width=0.25\textwidth]{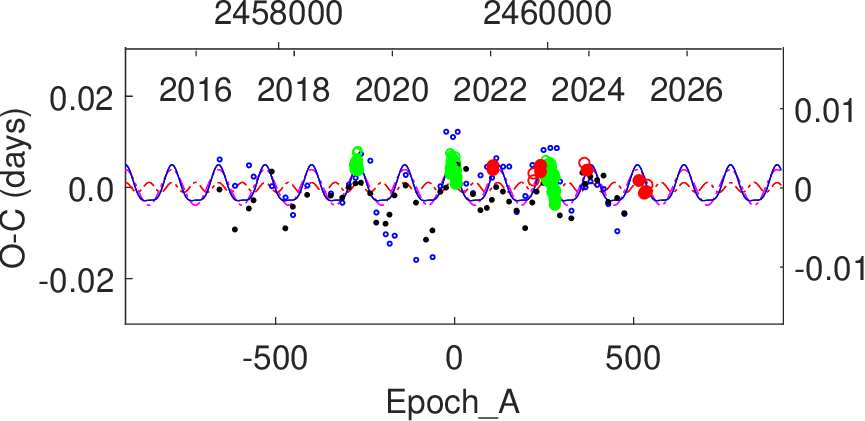}}
  \put(120,70){\includegraphics[width=0.25\textwidth]{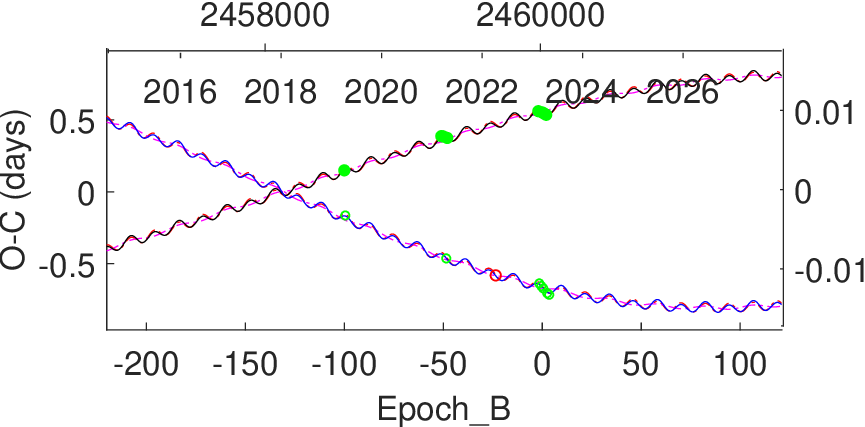}}
  \put(120,1){\includegraphics[width=0.25\textwidth]{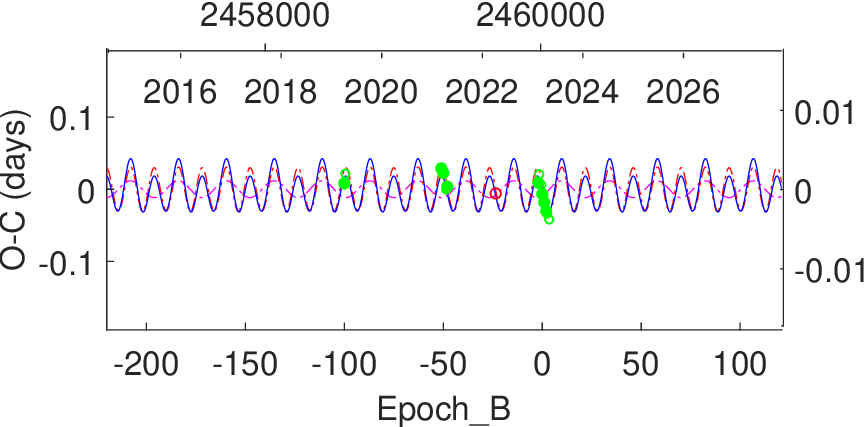}}
   \put(52,146){ {\textsf{Pair A}}}
   \put(50,56){ {\textsf{Pair B}}}
   \put(205,158){ {\textsf{Pair A}}}
   \put(205,102){ {\textsf{Pair B}}}
   \put(205,20){ {\textsf{Pair B}}}
  \end{picture}
  \caption{Analysis of the ASASSN-V J091951.17-593306.9 system. In the right-hand side figures are the ETV variations plotted twice for pair B. The upper plot is the complete ETV diagram (i.e. apsidal motion + R\"omer delay + physical delay). The lower figure is plotted after subtraction of the apsidal motion. The blue curve is the final combined fit, the red one stands for the physical delay, and the magenta is the R\"omer delay (LITE).}
  \label{Fig_asas091951}
 \end{figure}

As for the previous cases, we used the TESS photometry for detailed modelling of both inner pairs, A and B. The results are given in Table \ref{TabLC}, while the fit is plotted in Fig. \ref{Fig_asas091951}. It is evident that both pairs A and B are well detached. Both pairs have a similar brightness, as can be seen from the similar values of the third lights for both fittings. Pair B has a rather eccentric orbit and apsidal motion ($e=0.17$, $U=37$~yr). There is still a possibility that pair A also has a very small eccentricity, but only at a level of about 0.001. This is still barely detectable, even with precise TESS data.

For this system, we had to use a combined approach of fitting LITE (R\"omer delay) together with the physical delay (i.e. both equations 1 + 2 above) due to the very short mutual orbit. The two doubles revolve around a common barycenter with a periodicity of only about 347 days (the ratio of the inner period, $P_B$, over $P_{AB}$ is only 24). With such an approach, instead of deriving five LITE parameters ($p_{AB}$, $e$, $\omega$, $A$, and $T_0$) we have to also include another three ($i_m$, $v_m$, and $M_A/M_B$) in our computations. This means that the dynamical interaction between the orbits is not negligible at all. One can even expect some interaction of the orbits, such as the orbital precession of either of the orbits in the case of a non-coplanar configuration of the system (see e.g. \citealt{2022MNRAS.515.3773B}). This should lead to eclipse depth variations over a longer timescale.

 \subsection{WISE J100820.07-731554.0}

The star WISE J100820.07-731554.0 (= TIC 453459782, $V=14.9$ mag out of eclipse) is the fourth system studied in our sample, and one of the faintest. It consists of two quite dissimilar binaries. The more prominent one is a contact-like binary with an orbital period of about 0.37~days, discovered as an EB even in WISE data \citep{2018ApJS..237...28C}. The other EB is well detached, with a period of about 7.9~days. Moreover, the system was also resolved as a double by GAIA \citep{2023A&A...674A...1G}. The two components are about 0.8$^{\prime\prime}$ distant and are probably bound together, since both share a similar parallax and proper motion.

  \begin{figure}
 \centering
 \begin{picture}(380,145)
  \put(1,75){\includegraphics[width=0.22\textwidth]{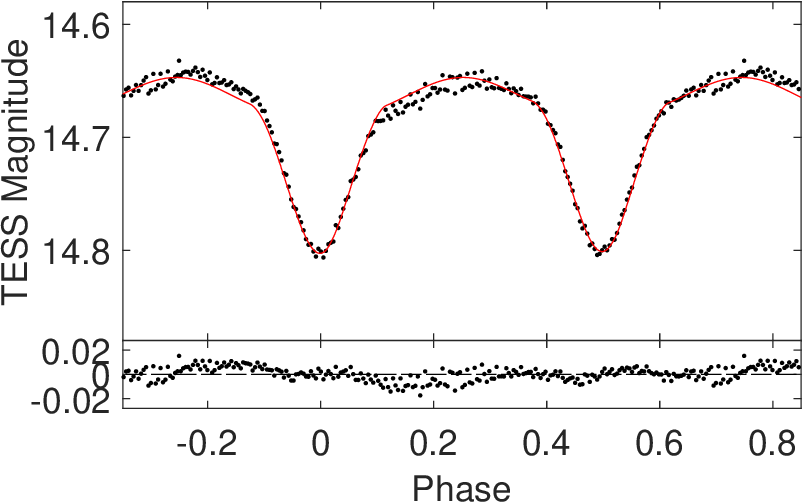}}
  \put(1,0){\includegraphics[width=0.22\textwidth]{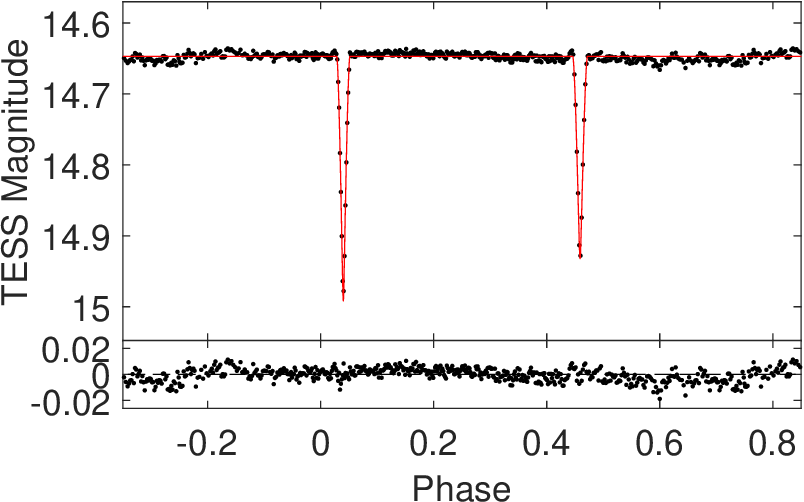}}
  \put(120,79){\includegraphics[width=0.25\textwidth]{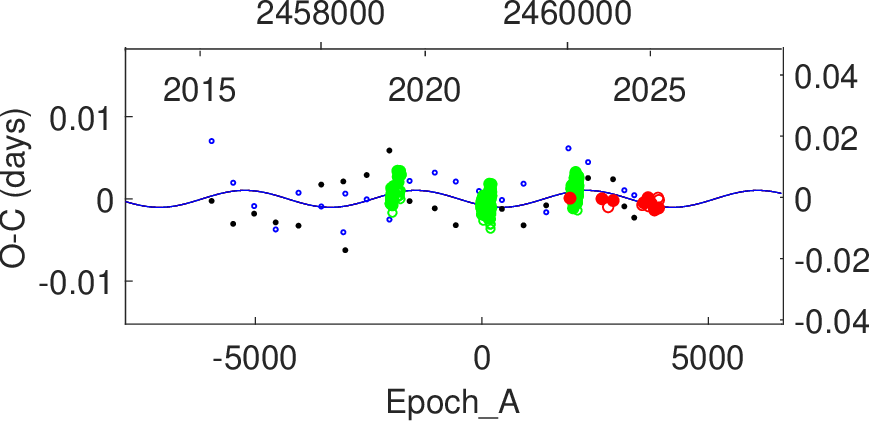}}
  \put(120,3){\includegraphics[width=0.25\textwidth]{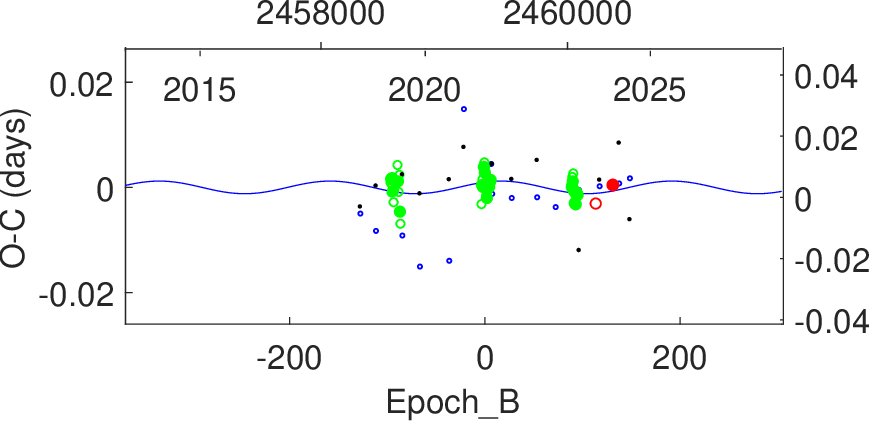}}
   \put(52,102){ {\textsf{Pair A}}}
   \put(52,28){ {\textsf{Pair B}}}
   \put(205,97){ {\textsf{Pair A}}}
   \put(205,21){ {\textsf{Pair B}}}
  \end{picture}
  \caption{Analysis of the WISE J100820.07-731554.0 system. The description of the plots and colour coding for individual symbols are the same as in Figure \ref{Fig_Czev1254}. The ETV diagram of pair B was plotted after subtraction of the slow apsidal motion.}
  \label{Fig_wise}
 \end{figure}
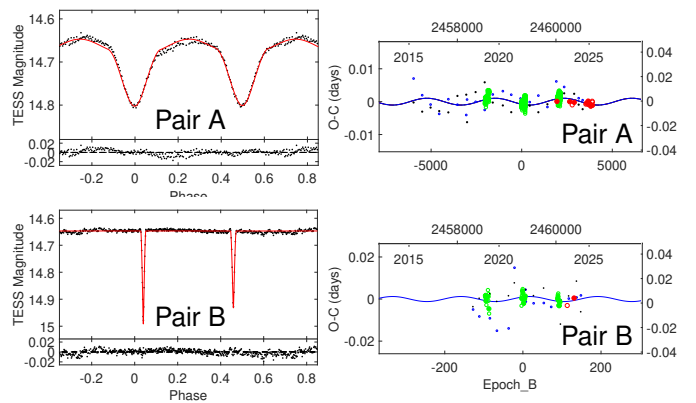

A similar approach as in previous cases was also used here. The LC fitting was based on the precise TESS data (from sector 65). The results of this fitting are presented in Table \ref{TabLC}. The final fit of the inner pairs, A and B, is plotted in Figure \ref{Fig_wise}. One can see that both pairs have a quite comparable brightness (and hence probably also masses), pair A being perhaps slightly more prominent. There is a slight asymmetry in the LC of pair A that is especially visible in regions outside of eclipses, probably due to spots on the surface of either of the stars.

Our ETV analysis was performed with the collection of all available data. However, the variation is only very small, at the edge of detectability, even for the precise TESS data. Pair B shows very long apsidal motion in its eccentric orbit ($e=0.14$, $U>1000$~yr). The results of our fitting are plotted in Figure \ref{Fig_wise}, and the parameters as well as ephemerides for both pair A and pair B are given in Table \ref{TabETV}. Due to the low quality of the data and low amplitude of the ETV variation, we only used a simplified approach of a circular A-B orbit. The periodicity is barely visible, and needs to be confirmed with new dedicated observations of reasonable quality in the coming years, if possible. The predicted 7~mas separation of the two A\&B binaries that results from our analysis indicates that this pair cannot be easily attributed to the detected components from GAIA (about 100$\times$ larger separation). This gives us a weak indication of its quintuple nature.

 \subsection{ASASSN-V J143536.01-721459.4}

The next system, ASASSN-V J143536.01-721459.4 (= TIC 401924368, $V=12.71$ mag out of eclipse), is the brightest star in our sample. It was originally detected as an EB in the ASAS-SN data \citep{2018MNRAS.477.3145J} and later classified as doubly eclipsing by \cite{2022A&A...664A..96Z}. Both of its inner binaries are detached with a rather long orbital periods. Pair A shows quite deep eclipses (about 0.3~mag), while pair B is about ten times shallower.

\begin{figure}
 \centering
 \begin{picture}(380,145)
  \put(1,75){\includegraphics[width=0.22\textwidth]{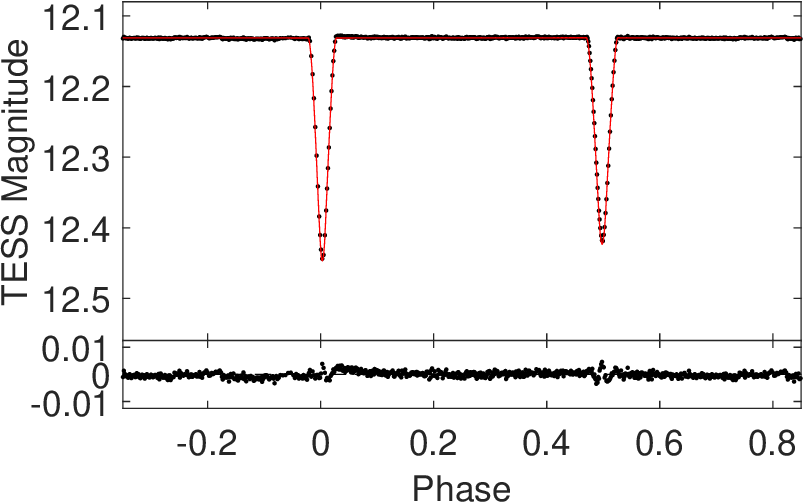}}
  \put(1,0){\includegraphics[width=0.22\textwidth]{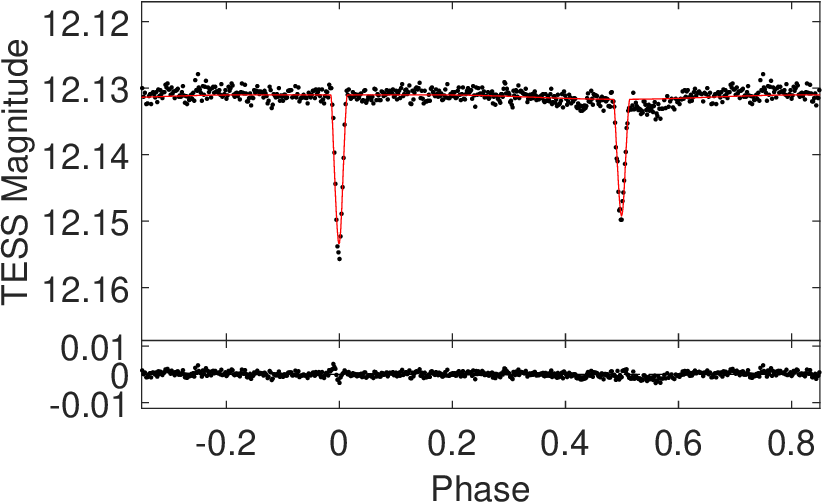}}
  \put(120,80){\includegraphics[width=0.25\textwidth]{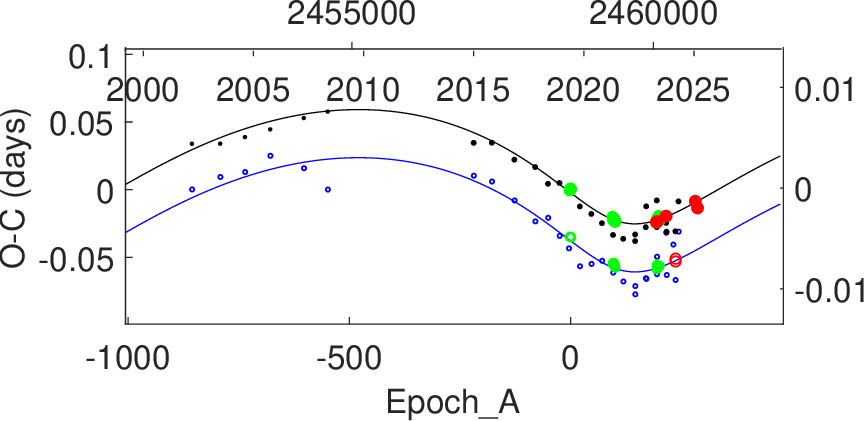}}
  \put(120,4){\includegraphics[width=0.25\textwidth]{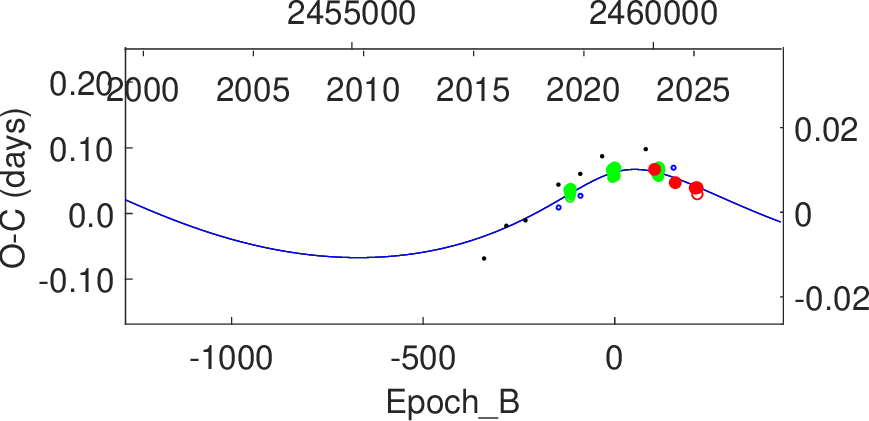}}
   \put(52,102){ {\textsf{Pair A}}}
   \put(52,26){ {\textsf{Pair B}}}
   \put(155,98){ {\textsf{Pair A}}}
   \put(205,22){ {\textsf{Pair B}}}
  \end{picture}
  \caption{Analysis of the ASASSN-V J143536.01-721459.4 system. The description of the plots and colour coding for individual symbols are the same as in Figure \ref{Fig_Czev1254}. }
  \label{Fig_asas143536}
 \end{figure}
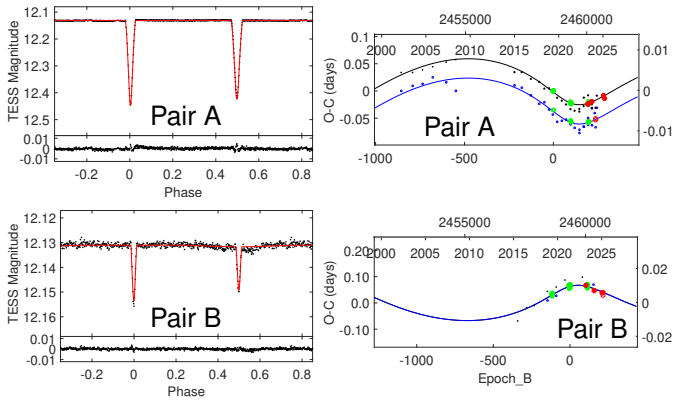

No detailed analysis for this system exists; therefore, we analysed the TESS data and performed the LC fitting. Using sector 65 we arrived at the LC parameters given in Table \ref{TabLC}. The final plots for both inner pairs are given in Fig.\ref{Fig_asas143536}. The two pairs have rather different brightnesses, with pair A dominating significantly. It is therefore not surprising that the ETV analysis presented in Figure \ref{Fig_asas143536} also shows that pair B has a higher amplitude of the ETV variation. It is still not a very robust solution due to the poor quality of the eclipsing times data for pair B, and its amplitude could be even higher. This is another system for which new data in the coming years would be beneficial to finally confirm our hypothesis with a higher conclusiveness.

 \subsection{Corot 310284765}

The last system in our set of analysed stars is also the faintest one, named Corot 310284765 (= TIC 44874302, $V=15.28$ mag out of eclipse). The doubly eclipsing nature was first detected by \cite{2017MNRAS.471.1230H} using the Corot data. However, since then no detailed analysis of the star has been published to date. According to its spectral classification (see Table \ref{systemsInfo}) this is the system with the earliest spectral type; hence, it also has the highest temperature in our sample. According to GAIA DR3, there are also some close-by components about 2.7$^{\prime\prime}$ and 4.3$^{\prime\prime}$ away from Corot 310284765. Nonetheless, it is unlikely that these objects are gravitationally bound to the star, as is evidenced by their distinct parallaxes and proper motions.

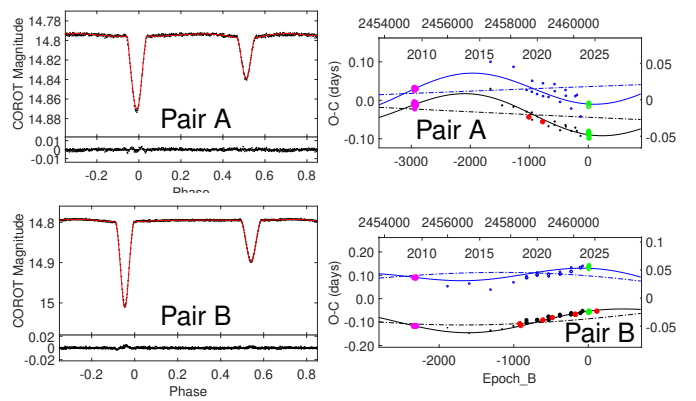
\begin{figure}
 \centering
 \begin{picture}(380,145)
  \put(1,75){\includegraphics[width=0.22\textwidth]{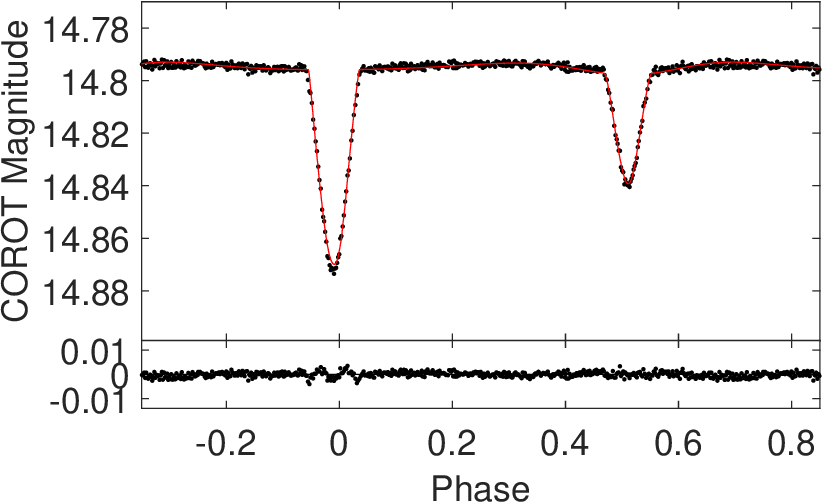}}
  \put(1,0){\includegraphics[width=0.22\textwidth]{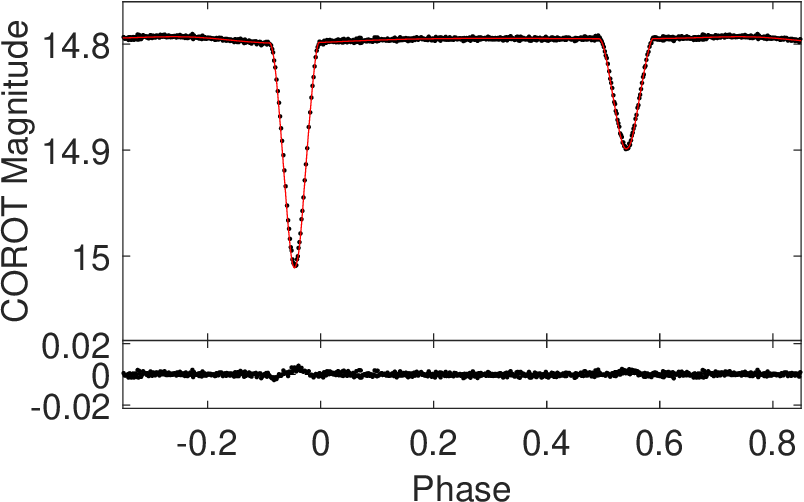}}
  \put(120,79){\includegraphics[width=0.25\textwidth]{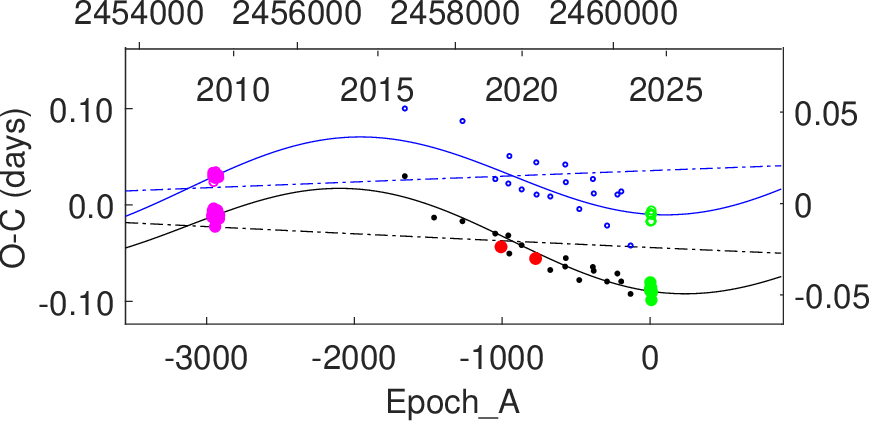}}
  \put(120,4){\includegraphics[width=0.25\textwidth]{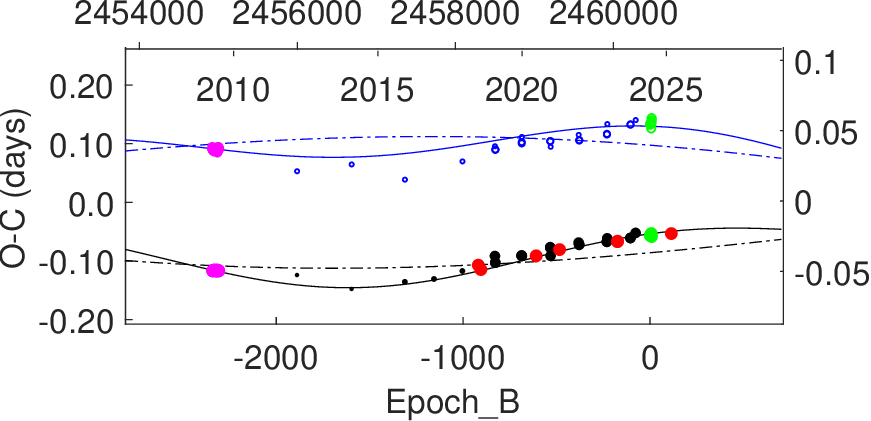}}
   \put(54,101){ {\textsf{Pair A}}}
   \put(54,27){ {\textsf{Pair B}}}
   \put(150,97){ {\textsf{Pair A}}}
   \put(206,20.5){ {\textsf{Pair B}}}
  \end{picture}
  \caption{Analysis of the Corot 310284765 system. The description of the plots and colour coding for individual symbols are the same as in Figure \ref{Fig_Czev1254}. The Corot data are shown as magenta symbols. }
  \label{Fig_corot}
 \end{figure}

This is the only system from our sample for which we used the Corot photometric data for LC modelling instead of TESS. This is due to their much better quality (the telescope has a significantly larger aperture, and hence the data obviously have a much lower scatter). Owing to the better angular resolution of the Corot data (about 10$\times$ better compared to TESS), also the two EBs appear to be deeper (approximately 2$\times$) in the Corot data compared to the TESS data. The results are given again in our Figure \ref{Fig_corot}, where one can see the fits for both pairs, and the final parameters are given in Table \ref{TabLC}.

Both pair A and pair B show slightly eccentric orbits, which is clearly seen in the LCs. This can also be seen in our ETV analysis (plotted in Figure \ref{Fig_corot}), in which the slow apsidal motion is evident for both pairs. Apart from that,  periodic variation in the pairs as a consequence of movement around a common barycenter is also visible. The solution is still rather preliminary, since its coverage is poor, and new observations can still shift the solution significantly. Nevertheless, the bound nature is proven with our presented solution.

\begin{table*}
\caption{Derived parameters for the two inner binaries, A and B.}
 \label{TabLC}
 \tiny
  \centering \scalebox{0.82}{
\begin{tabular}{ c | c | c | c | c | c | c }
   \hline\hline
 \multicolumn{1}{c|}{System } &    CzeV1254          &ASASSN-V J070838.27-171952.9&ASASSN-V J091951.17-593306.9 & WISE J100820.07-731554.0 &  ASASSN-V J143536.01-721459.4 & Corot 310284765  \\ \hline
  \multicolumn{1}{c|}{ }      &  \multicolumn{6}{c}{{\sc p a i r \,\, A}} \\ \hline
  $i$ [deg]                   &  74.14 $\pm$ 0.48     &     84.54 $\pm$ 0.22      &       88.79 $\pm$ 0.31      &    71.51 $\pm$ 0.59     &  87.78 $\pm$ 0.04  &  77.66 $\pm$ 0.11   \\
 $q=\frac{M_2}{M_1}$          &  1.00 $\pm$ 0.03      &      1.00 (fixed)         &       1.00 (fixed)          &    1.03 $\pm$ 0.05      &    1.00 (fixed)    &   1.00 (fixed)      \\
 $e$                          &     0                 &          0                &              0              &           0             &   0.06 $\pm$ 0.01  &   0.12 $\pm$ 0.02   \\
 $T_1$ [K]                    &  5731 (fixed)         &      9390 (fixed)         &       5783 (fixed)          &     5143 (fixed)        &    5796 (fixed)    &   20000 (fixed)     \\
 $T_2$ [K]                    &  5504 $\pm$  74       &      9079 $\pm$ 88        &        5391 $\pm$ 45        &     5124 $\pm$ 89       &    5740 $\pm$ 51   &  13940 $\pm$ 72     \\
 $R_1/a$                      &  0.364 $\pm$ 0.002    &     0.126 $\pm$ 0.002     &      0.114 $\pm$ 0.002      &   0.352 $\pm$ 0.004     &  0.084 $\pm$ 0.002 &  0.191 $\pm$ 0.004  \\
 $R_2/a$                      &  0.255 $\pm$ 0.003    &     0.117 $\pm$ 0.002     &      0.102 $\pm$ 0.002      &   0.354 $\pm$ 0.005     &  0.081 $\pm$ 0.002 &  0.154 $\pm$ 0.003  \\
 $L_1$ [\%]                   &  35.4 $\pm$ 0.5       &      28.8 $\pm$ 0.4       &       31.5 $\pm$ 0.5        &   23.7 $\pm$ 0.4        &   35.4 $\pm$ 0.6   &  39.6 $\pm$ 0.8     \\
 $L_2$ [\%]                   &  13.1 $\pm$ 0.6       &      22.2 $\pm$ 0.4       &       18.6 $\pm$ 0.3        &   23.0 $\pm$ 0.5        &   32.0 $\pm$ 0.5   &   8.9 $\pm$ 0.6     \\
 $L_3$ [\%]                   &  51.5 $\pm$ 1.4       &      49.0 $\pm$ 1.2       &       50.9 $\pm$ 1.8        &   53.3 $\pm$ 2.3        &   32.6 $\pm$ 1.1   &  51.5 $\pm$ 1.2     \\ \hline
  \multicolumn{1}{c|}{ }      & \multicolumn{6}{c}{{\sc p a i r \,\, B}}\\ \hline
 $i$ [deg]                    &   61.75 $\pm$ 0.67    &      82.24 $\pm$ 0.27     &       89.41 $\pm$ 0.18      &    89.69 $\pm$ 0.13     &  84.01 $\pm$ 0.10  &  83.33 $\pm$ 0.09   \\
 $q=\frac{M_2}{M_1}$          &  1.01 $\pm$ 0.02      &      1.00 (fixed)         &        1.00 (fixed)         &     1.00 (fixed)        &   1.00 (fixed)     &   1.00 (fixed)      \\
 $e$                          &     0                 &          0                &       0.17 $\pm$ 0.02       &     0.14 $\pm$ 0.02     &        0           &  0.15 $\pm$ 0.02    \\
 $T_1$ [K]                    &  5731 (fixed)         &      9390 (fixed)         &       5783 (fixed)          &     5143 (fixed)        &    5796 (fixed)    &   20000 (fixed)     \\
 $T_2$ [K]                    &  5451 $\pm$ 52        &      8352 $\pm$ 97        &        5802 $\pm$ 40        &     5059 $\pm$ 62       &    5684 $\pm$ 78   &  13967 $\pm$ 61     \\
 $R_1/a$                      &  0.379 $\pm$ 0.003    &     0.114 $\pm$ 0.003     &      0.033 $\pm$ 0.001      &    0.039 $\pm$ 0.001    &  0.069 $\pm$ 0.002 &  0.152 $\pm$ 0.002  \\
 $R_2/a$                      &  0.380 $\pm$ 0.003    &     0.102 $\pm$ 0.004     &      0.036 $\pm$ 0.001      &    0.039 $\pm$ 0.001    &  0.067 $\pm$ 0.002 &  0.149 $\pm$ 0.002  \\
 $L_1$ [\%]                   &  28.6 $\pm$ 0.2       &     19.3 $\pm$ 0.7        &        24.6 $\pm$ 0.6       &    28.8 $\pm$ 0.6       &  16.0 $\pm$ 0.5    &  32.2 $\pm$ 0.6     \\
 $L_2$ [\%]                   &  22.0 $\pm$ 0.2       &     10.0 $\pm$ 0.6        &        28.3 $\pm$ 0.5       &    25.5 $\pm$ 0.4       &  12.7 $\pm$ 0.5    &  17.4 $\pm$ 0.5     \\
 $L_3$ [\%]                   &  49.4 $\pm$ 1.0       &     70.7 $\pm$ 1.8        &        48.1 $\pm$ 1.1       &    45.7 $\pm$ 1.2       &  71.3 $\pm$ 1.7    &  50.4 $\pm$ 0.9     \\
 \noalign{\smallskip}\hline
\end{tabular}} \\
\end{table*}

\begin{table*}
\caption{Results of the combined analysis of the ETV data for both pair A and pair B.}
 \label{TabETV}
 \tiny
  \centering \scalebox{0.73}{
\begin{tabular}{c | c c c c c c }
   \hline\hline
 \multicolumn{1}{c|}{    } &   CzeV1254                & ASASSN-V J070838.27-171952.9 & ASASSN-V J091951.17-593306.9 & WISE J100820.07-731554.0  & ASASSN-V J143536.01-721459.4 & Corot 310284765    \\ \hline
  $JD_{0,A}$ [HJD-2450000] & 8827.351 $\pm$ 0.002      &  7042.819 $\pm$ 0.003        &    9314.277 $\pm$ 0.001      &  9312.384 $\pm$ 0.001     & 8642.166 $\pm$ 0.004     &  9917.570 $\pm$ 0.005  \\
 $P_A$ [day]               & 0.7153484 $\pm$ 0.0000006 & 4.3004974 $\pm$ 0.0000042    &   2.6686745 $\pm$ 0.0000012  & 0.3683154 $\pm$ 0.0000001 & 7.353842 $\pm$ 0.0000014 &  1.8754183 $\pm$ 0.0000015 \\
  $JD_{0,B}$ [HJD-2450000] & 8827.747 $\pm$ 0.002      &  9211.742 $\pm$ 0.002        &   10018.034 $\pm$ 0.027      &  9338.383 $\pm$ 0.020     & 9379.268 $\pm$ 0.008     &  9980.211 $\pm$ 0.011   \\
 $P_B$ [day]               & 0.3625933 $\pm$ 0.0000003 & 3.7672118 $\pm$ 0.0000075    &   14.342218 $\pm$ 0.000408   & 7.9453385 $\pm$ 0.0000348 & 6.3685673 $\pm$ 0.000009 &  2.3711257 $\pm$ 0.000019 \\
 $p_{AB}$ [yr]             &  15.6 $\pm$ 0.8           &      3.91 $\pm$ 0.17         &     0.95 $\pm$ 0.25          &  3.81 $\pm$ 0.60          &  27.4 $\pm$ 1.4          &  22.6 $\pm$ 3.5          \\
 $A_A$ [d]                 & 0.027 $\pm$ 0.002         &     0.008 $\pm$ 0.01         &     0.004 $\pm$ 0.001        &  0.001 $\pm$ 0.001        & 0.042 $\pm$ 0.004        &  0.047 $\pm$ 0.009       \\
 $A_B$ [d]                 & 0.026 $\pm$ 0.002         &     0.015 $\pm$ 0.03         &     0.012 $\pm$ 0.003        &  0.001 $\pm$ 0.001        & 0.067 $\pm$ 0.006        &  0.034 $\pm$ 0.011       \\
 $e$                       & 0.791 $\pm$ 0.014         &       0.0 (fixed)            &     0.011 $\pm$ 0.005        &     0.0 (fixed)           &  0.421 $\pm$ 0.020       &  0.007 $\pm$ 0.002       \\
 $\omega$ [deg]            & 295.5 $\pm$ 7.9           &          --                  &      1.9 $\pm$ 9.5           &         --                &  250.6 $\pm$ 16.0        &  165.9 $\pm$ 12.0        \\
 $T_0$ [HJD-2450000]       & 5226.5 $\pm$ 244.5        &    9316.5 $\pm$ 119.0        &    9552.798 $\pm$ 117.0      &   7389 $\pm$ 253          &  9424.1 $\pm$ 456.1      &  8417 $\pm$ 905          \\
 $i_m$  [deg]              &     --                    &          --                  &      69.3 $\pm$ 6.1          &        --                 &        --                &        --                \\
 $v_m$  [deg]              &     --                    &          --                  &      42.7 $\pm$ 7.3          &        --                 &        --                &        --                \\
 $M_A/M_B$                 &     --                    &          --                  &      0.51 $\pm$ 0.04         &        --                 &        --                &        --                \\
 \noalign{\smallskip}\hline
\end{tabular}} \\
\end{table*}

\section{Conclusions}  \label{Conclusions}

The doubly eclipsing quadruple systems still present a challenge when it comes to modelling their observational data. These are usually only photometric observations, because of their lower brightness. To date, only a select number of doubly eclipsing systems have been studied in detail, combining both photometry and spectroscopy (see e.g. \citealt{2021ApJ...917...93K}, or \citealt{2023MNRAS.524.4220P}). These systems provide us with a unique insight into the star formation and consequent orbital evolution due to the possible derivation of overall architecture of the whole quadruple, derivation of mass distribution, and mutual inclination of the orbits, for example.

Quite recently \citep{2023A&A...675A.113Z}, we also discussed the possibility of mean motion resonances in the quadruple system (i.e. the A and B orbital periods) and the spectral type of the component stars. However, here in our sample the individual systems seem to be rather distant from the resonances. The closest ones are CzeV1254, which is about 1.4\% distant from the 2:1 resonance, and Corot 310284765 about 1.1\% distant from the 4:5 resonance. However, both values are too far away from any potential resonance configuration.

\begin{figure}
 \centering
 \begin{picture}(380,320)
  \put(30,160){\includegraphics[width=0.38\textwidth]{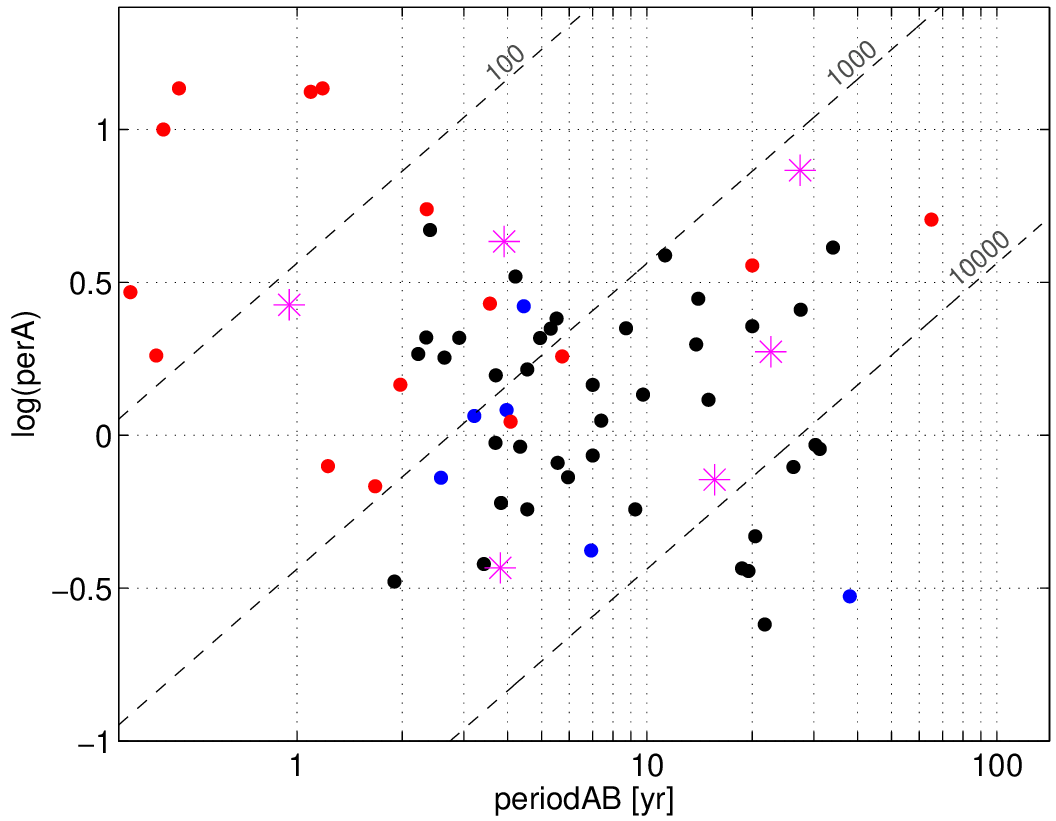}}
  \put(30,0){\includegraphics[width=0.38\textwidth]{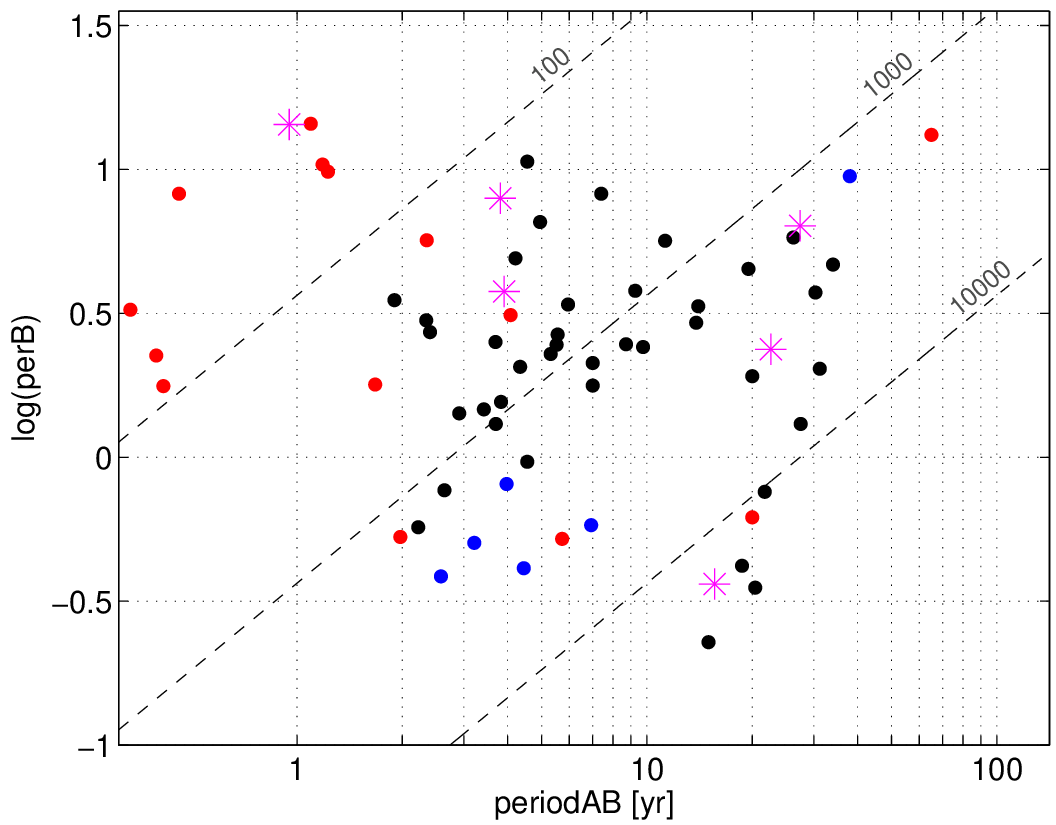}}
  \end{picture}
   \caption{Inner and outer periods for the 2+2 quadruples. Individual proven 2+2 systems: red dots published by T. Borkovits group, black dots by our group, and the blue ones are the rest of other authors. The newly presented systems in this paper are denoted by magenta asterisk signs. The sloped dashed lines denote the period ratios 100, 1000, and 10000, respectively. One can clearly see that the red dots have in general a shorter A-B period than our systems plotted in black.}
  \label{statOrbits}
 \end{figure}

The two figures plotted in Fig. \ref{statOrbits} show the proven 2+2 doubly eclipsing quadruples. They are compared to the sets of systems published by the group around T. Borkovits (using their photodynamical modelling), our group, and the other systems published by various other authors (unconnected to any previous groups of authors, see e.g. \citealt{2023A&A...674A.170A}). As one can see, the systems discovered and analysed by T. Borkovits et al. are usually more compact (i.e. there is a smaller outer-to-inner period ratio), and sometimes have a mutual A-B period shorter than 100$\times$ the inner period. Only one of our systems is like that (here presented ASASSN-V J091951.17-593306.9), which is the reason why we had to use both the geometrical LITE as well as the dynamical term in our calculations for this particular system.

In general, we can say that the presented method of ETV detection and discovering multiples with periods of several years provides us with an excellent method of filling the incomplete statistics. When examining the distribution of periods as presented in \cite{Tok2008}, one can see that the longer periods are usually discovered by astrometry (interferometry, CPM pairs, etc.), while the shorter ones are discovered by spectroscopy. The gap in between can be effectively filled with our presented ETV method, which preferably finds systems with periods of years to decades.

We also discussed the possibility of interferometric detections of the doubles based on the period of the A-B pair and the distance to the system. However, for most of the systems this is impractical because of their lower brightness and only a small angular separation. On the other hand, for the system CzeV1254, there is other indirect evidence that the system is not a single star, even in the GAIA DR3 catalogue. It provides us with information about the so-called ‘GAIA re-normalised unit weight error’ (RUWE) value, which indicates that this particular star also shows some non-negligible movement (RUWE exceeding 2.5). We can only speculate that this is caused by our detected 15.6-yr orbit. Future observations over the coming years will shed light on the precise orbital parameters for all of the systems, and are therefore desirable.

\section{Data availability}

Table \ref{CDSdata} is only available in electronic form at the CDS via anonymous ftp to cdsarc.u-strasbg.fr (130.79.128.5) or via http://cdsweb.u-strasbg.fr/…+A/. 

\medskip

\begin{acknowledgements}
An anonymous referee is greatly acknowledged for his/her helpful and critical suggestions. We do thank the ZTF, ASAS-SN, SWASP, Atlas, OGLE, and TESS teams for making all of the observations easily public available. We are also grateful to the ESO team at the La Silla Observatory for their help in maintaining and operating the Danish 1.54m telescope. The research of P.Z., J.K., and J.M. was also supported by the project {\sc Cooperatio - Physics} of Charles University in Prague. The research of JM was supported by the Czech Science Foundation (GACR) project no. 24-10608O. We would like to thank the Pierre Auger Collaboration for the use of its facilities. This work is supported by MEYS (Czech Republic) under the projects Czech Republic MEYS LM2023032, LM2023047, and CZ.02.01.01/00/22$\_$008/0004632. The observations by ZH in Velt\v{e}\v{z}e were obtained with a CCD camera kindly borrowed by the Variable Star and Exoplanet Section of the Czech Astronomical Society. This research made use of Lightkurve, a Python package for TESS data analysis \citep{2018ascl.soft12013L}. This research has made use of the SIMBAD and VIZIER databases, operated at CDS, Strasbourg, France and of NASA Astrophysics Data System Bibliographic Services. This work has made use of data from the European Space Agency (ESA) mission {\it Gaia} (\url{https://www.cosmos.esa.int/gaia}), processed by the {\it Gaia} Data Processing and Analysis Consortium (DPAC, \url{https://www.cosmos.esa.int/web/gaia/dpac/consortium}). Funding for the DPAC has been provided by national institutions, in particular the institutions participating in the {\it Gaia} Multilateral Agreement.
\end{acknowledgements}

\begin{appendix}

\section{Online data}

\begin{table}[h!]
  \caption{Ground-based photometry of the analysed systems.}  \label{CDSdata}
  \begin{tabular}{c c c c c}\\[-3mm]
  \hline\\[-3mm]
  Target name & HJD-2400000 & Diff.mag.  & Filter &  Observer  \\ \hline
   CzeV1254 & 59566.51944 & 0.269 & R & L.C. \\
   CzeV1254 & 59566.52085 & 0.243 & R & L.C. \\
   CzeV1254 & 59566.52225 & 0.273 & R & L.C. \\ 
   CzeV1254 & 59566.52366 & 0.224 & R & L.C. \\
  \hline
\end{tabular}\\[1mm]
{\small Note: This table is available in its entirety in machine-readable form in electronic version of the journal, and via a CDS. A portion is shown here for guidance regarding its form and content. }
\end{table}

\end{appendix}

\end{document}